\documentclass[12pt,aps,prb,preprint]{revtex4}   % style for Physical Review B and AJP are similar

\usepackage{amsmath}    % need for subequations
\usepackage{amsfonts}  %note how statements can be commented out
\usepackage{amssymb}
\usepackage{graphicx}   % for figures
\usepackage{here}
\draft
\begin{document}

\title{The Equivalence Principle as a Stepping Stone from Special to General Relativity: A Socratic Dialog}
%Lines break automatically or can be forced with \\
\author{S. P. Drake}
\affiliation{School of Chemistry \& Physics\\
 The University Of Adelaide \\Adelaide
 SA 5005 AUSTRALIA}
\email{samuel.drake@adelaide.edu.au}   %optional

\begin{abstract}
In this paper we show how the student can be led to an understanding
of the connection between special relativity and general relativity by
considering the time dilation effect of clocks placed on the surface
of the Earth. This paper is written as a Socratic dialog between a
lecturer Sam and a student Kim.
\end{abstract}

\maketitle

\section{Setting the Scene}
Sam is in the office and has just finished reading Plato's {\it
Meno}~\cite{plato} in which Socrates uses a self-discovery
technique to teach a boy Pythagoras' theorem. Sam is inspired by
this dialog and is pondering its applicability to lecturing
undergraduate physics when a tap on the door breaks that chain of
thought. Kim enters the room looking bleary eyed and pale. ``Been
out celebrating the last lecture of the year'' Sam surmises,
little knowing that other things have kept Kim awake.

\section{The Dialog}
\begin{itemize}
\item[Kim:] Your lectures on special relativity fascinated me, and
when I got home I wondered if I could construct a simple experiment to
prove or disprove time dilation, the aspect of special relativity that
interests me the most. While lying in bed before dozing off I realised
that a clock placed at the equator should run slower than a clock
placed at the pole. So I did a little calculation and found that
special relativity predicts that a clock on the equator runs slower by
about 100 nanoseconds per day with respect to a clock at the
pole. While this effect is not large it is certainly measurable with
modern atomic clocks. So I went onto the internet to see if I could
find any reference to such an experiment and to my surprise I
couldn't.

I was starting to get so frustrated that I couldn't sleep. I glanced
at the clock (3am). I thought to myself ``How accurate is my clock?
I should check it against internet time.'' Then it occurred to me
that the world timing standard organisations must mention a latitude
effect on local clock accuracies. So I got onto the internet again
and checked \textsl{The Bureau International des Poids et Mesures
(BIPM)}\cite{bipm} as they calculate the international atomic time
(TAI). BIPM calculate TAI from atomic clocks located in more than 30
countries around the world. I was sure that I must find something
about the latitude effect on their web site. After spending hours
trawling through the site and then other sites on the web, I came up
with nothing. There was a discussion of the relativistic effect of
placing clocks at high altitudes, but nothing about latitude. In my
despair I gave up and collapsed into a fitful sleep.

I came to see you today in the hope that you could cure my
insomnia.

\item[Sam:] You are in good company in thinking that clocks at the
equator and the pole should tick at different rates. Einstein
himself predicted as much in his famous 1905 paper on the special
theory of relativity.\cite{harvey05} Luckily for physics the effect
was not measurable with the instruments of the day as Einstein's
prediction would have failed to match experiment.

Let us return to your findings:
\begin{enumerate}
\item According to the special theory of relativity a clock
located at the equator should run slower than one at the pole
\item All clocks located at sea-level on the Earth's surface tick
at the same rate, regardless of latitude
\end{enumerate}
To help you understand how both apparently contradictory
statements can be true I will ask you a question.

If the Earth was a rotating perfect fluid and we could ignore the
gravitational effects of the Sun and the Moon what shape would it
be?

\item[Kim:] Well, I don't see how this is relevant, but I would answer
your question by drawing a free-body diagram. Can I use your
blackboard?
%
%INSERT FIGURE 1 HERE
\begin{center}
\begin{figure}
\label{fig:fbd} \includegraphics[width=.8\textwidth]{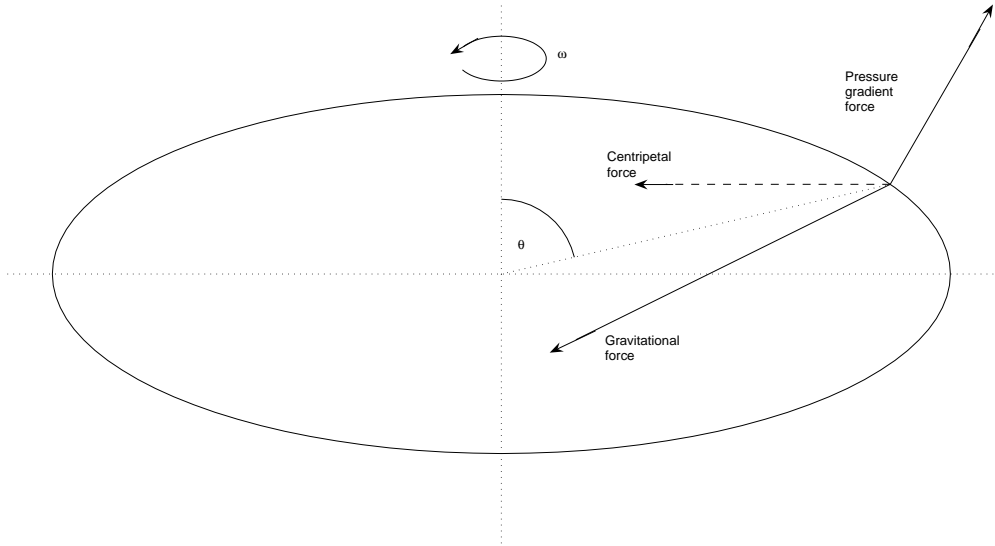}
 \caption{Free body diagram for a mass placed on the surface of the
 Earth as seen from space}
\end{figure}
\end{center}
Now let me see \ldots consider a test mass placed on the surface of
the Earth. We know that the forces acting on the test mass are the
outward pointing force due to the difference in pressure and the
inward pointing force due to gravity. If the test mass is in
hydrostatic equilibrium then the pressure force must be
perpendicular to the surface and the sum of the gravitational and
pressure forces is the centripetal force, which is perpendicular to
the axis of rotation.\cite{bruton04} Hmmm\ldots you would have a
complicated integral equation to solve because the direction of the
gravitational force vector would depend on the distribution of mass,
furthermore the pressure gradient would be perpendicular to the
surface we are trying to calculate. It seems to be a complicated
problem and, to be honest, I am not sure that I could solve it.

\item[Sam:] It is a difficult problem and one whose solutions
involve hyperbolic and elliptic functions. Chandrasekhar has devoted
a whole book to the subject.\cite{cha87} Before we travel that
arduous mathematical road let us see if we can use some physics to
help us. Taking our model of Earth as a rotating perfect fluid, is
the Earth an equipotential surface?

\item[Kim:] (\textsl{Thinks} \ldots) Yes.

\item[Sam:] Why?

\item[Kim:] Because if it wasn't the sea water would feel a force
$\vec{F}=-m\nabla \Phi$ and would move until $\nabla \Phi=0$
everywhere on the surface.

\item[Sam:] So if I told you what the Earth's gravitational field
is could you tell me the shape of the Earth?

\item[Kim:] Yes, I think I could.

\item[Sam:] How?

\item[Kim:] If you told me that the Earth's gravitational field is
$\Phi_{g}(r,\theta)$, where $r$ is the distance from the centre
and $\theta$ is the colatitude~\cite{colat} then I could calculate
the \textsl{effective} potential felt by an observer co-rotating
with the Earth by including the centripetal force:
\begin{equation}\label{eq:effPot}
\begin{array}{l}
    \Phi_{ep} = \Phi_{g}(r,\theta) - \frac12 \omega^2 r^2\sin^2\theta\\
     \theta\in[0,\pi] \\
     \mbox{$\theta = 0$ at the north pole, $\pi/2$ at equator and $\pi$ at the south pole}      \end{array}
\end{equation}
where $\omega$ is the Earth's rotation rate and $r$ is the distance
from the centre to the Earth's surface. The second term on the right
hand side of~Eq.\eqref{eq:effPot} is the so called ``centrifugal
potential''. Now we have already argued that a co-rotating observer
on the surface of the Earth feels no change in effective potential
regardless of their latitude, i.e., $\Phi_{ep}$ is constant.
Furthermore, you have told me that we know what the Earth's
gravitational field $\Phi_{g}(r,\theta)$ is, so all I need to do is
rearrange equation~Eq.\eqref{eq:effPot} and \textsl{voil\'{a}} we
have an expression for the shape of the Earth's surface. Mind you,
as $\Phi_g(r,\theta)$ may be a complicated function; I am not sure
that I can find an analytic expression for $r$ anyway.

All this is very interesting, but I don't see how it answers my
question about why clocks tick at the same rate on the Earth's
surface.

\item[Sam:] Patience, we are coming to that. First let us
investigate the discovery you have made, namely the shape of the
Earth. Let me see, I know I have it in here somewhere \ldots

\textsl{Sam flicks through some notes in the filing cabinet}

Ah here it is. Despite the Earth's complicated shape with mountains
and valleys its gravitational field can be modelled to a fractional
accuracy of $10^{-14}$ by:\cite{aud01}
\begin{equation}\label{eq:spacePot}
    \Phi_g(r,\theta) = \frac{-GM_e}{r} -\; \frac{J_2 G M_e a^2 (1-3\cos^2\theta)}{2r^3},
\end{equation}
where
\begin{itemize}
    \item[$\bullet$] $GM_e=3.98600442\times10^{14}$m$^3$s$^{-2}$ is the product of the gravitational constant and the mass of the
    Earth~\cite{mass}
    \item[$\bullet$] $J_2= 1.082636\times10^{-3}$ is a measure of the Earth's equatorial bulge and is related to the Legendre
    polynomials~\cite{bom86}
    \item[$\bullet$] $a=6378137$m is the Earth's equatorial radius~\cite{a}
\end{itemize}
To evaluate your equation for the Earth's surface (which incidentally is called the Geoid) you will
need an accurate value of the Earth's rotation
rate. \\
 \textsl{Sam shuffles through some files} \ldots \\
 Yes here it is~\cite{lid97}
\[\omega = 7.292116\times10^{-5}\mbox{rad$\;$s$^{-1}$}\quad.\]

Now your Geoid equation is going to be a bit tricky to solve
analytically so instead of doing that let us see if we are on the
right track. The easiest thing for us to do is to check that your
equation for the Earth's effective potential $\Phi_{ep}$ is the
same at the equator and the pole.
\begin{itemize}
    \item[$\bullet$] \underline{$\Phi_{ep}$ at the pole}: The Earth's mean polar radius is $\bar{c}=6356.76\pm0.07$km~\cite{lid97}
    \begin{eqnarray}\label{eq:effPotPole}
      \Phi_{ep}(r=\bar{c},\theta=0) &=& \frac{-GM_e}{\bar{c}} + \frac{J_2GM_ea^2}{\bar{c}^3} \nonumber \\
      & = & -6.2637\times 10^7m^2s^{-2}
    \end{eqnarray}
    \item[$\bullet$] \underline{$\Phi_{ep}$ at the equator}: The Earth's mean equatorial radius is $\bar{a}=6378.1\pm0.2$km~\cite{lid97}
\begin{eqnarray}\label{eq:effPotEq}
      \Phi_{ep}(r=\bar{a},\theta=pi/2) &=& \frac{-GM_e}{\bar{a}} - \frac{J_2GM_ea^2}{2\bar{a}^3} -\frac12\omega^2\bar{a}^2\nonumber \\
      & = & -6.2637\times 10^7m^2s^{-2}
    \end{eqnarray}
\end{itemize}
Look the two values for $\Phi_{ep}$ are the same!

What have you shown?

\item[Kim:]  We have shown that the Earth is indeed an
equipotential surface with respect to an observer sitting on the
surface. But Sam, this has nothing to do with the question I
originally asked you!

 \item[Sam:] Doesn't it? What did you ask me again?

\item[Kim:] I asked you why all clocks tick at the same rate on
the surface of the Earth when special relativity predicts that
they should run slower at the equator than at the pole.

\item[Sam:] Kim do you remember how we derived Einstein's famous
formula $E=mc^2$?

\item[Kim:] Yes, and to be honest I was a little disappointed with
it. Once we learnt that a constant speed of light lead to the
Lorentz transformations, the rest was just algebra.

\item[Sam:] Remind me of the algebra.

\item[Kim:] We got to the point that we realised that the proper
time interval, $d\tau$ must be defined as
\begin{equation}\label{eq:propTime}
    c^2d\tau^2 = c^2 dt^2 -d\vec{x}^2,
\end{equation}
with $dt$ and $d\vec{x}$ the coordinate time and space interval
respectively. Then we simply multiplied
equation~Eq.\eqref{eq:propTime} by $\frac{m^2c^2}{d\tau^2}$ to get
\begin{eqnarray}
  m^2c^4 &=& m^2c^4\left(\frac{dt}{d\tau}\right)^2-m^2\vec{u}^2c^2\quad\mbox{equating $\vec{u}$ with $\frac{d\vec{x}}{d\tau}$} \nonumber \\
  & = & m^2c^4\gamma^2-\vec{p}^2c^2\quad\mbox{since $\gamma = \frac{dt}{d\tau}$ and $\vec{p} = m\vec{u}$} \nonumber \\
  & = & E^2 -\vec{p}^2c^2\quad \mbox{since relativistic kinetic energy is $mc\gamma$}.
\end{eqnarray}
So if $\vec{p}=0$, then $E = mc^2$, like I said, just algebra.

\item[Sam:] Hmm, yes indeed. Suppose you are floating in a room with no windows or doors. All of a
sudden, you feel a force that throws you against the wall. If their were two possible forces,
gravitational or centrifugal, are you able to determine which force you are feeling?

\item[Kim:] I don't see how.

\item[Sam:] And what would you (sitting in this closed room) say
your time dilation was with respect to an observer who was not
feeling the centrifugal or gravitation force?

\item[Kim:] I think I see what you are getting at. I can't say whether the force is gravitational
or centrifugal, so I must treat their effects as the same. If I knew the force was centrifugal, I
would say that my time dilation with respect to a stationary observer depends only on my velocity
$v$, i.e., $\gamma= \frac{1}{\sqrt{1-v^2/c^2}}$.  As I don't know where the energy to thrust me
against the wall has come from, to be consistent, I must say that the time dilation depends only on
the effective potential, which is the sum of the gravitational and centripetal potentials.

\item[Sam:] Excellent! The idea that you can't know if the force
is a uniform gravitational force, or a combination of uniform
forces, is called the {\it equivalence principle}.~\cite{schutz85}
What does it tell you about clocks on the surface of the Earth?

\item[Kim:] Yes, yes, of course.  According to somebody standing anywhere on the surface of the
Earth, all their energy is effective potential energy $\Phi_{ep}$. The rate at which their clock
ticks depends only on this effective potential. We already showed that the effective potential over
the surface of the Earth is constant. So all clocks on the surface of the Earth tick at the same
rate. Eureka, I can sleep again!

\item[Sam:] Yes, you can sleep well indeed because you have just
discovered one of the fundamental arguments that led to the
development of the general theory of relativity. Before you go,
let me clarify one point. To determine the time dilation, you used
the effective potential which came from newtonian arguments about
gravitational and centrifugal forces. According to general
relativity the newtonian effective potential is an approximation
to the relativistic effective potential. This does not change your
conclusion in any way, the effective potential is still constant,
it just means that in general relativity we have a slightly
different version of $\Phi_{ep}$ (\textsl{see
Appendix~\ref{app:clockGR}}). Having said that, you should note
that for the Earth, the newtonian and relativistic effective
potentials are almost identical. To learn precisely what the
difference is, you will have to take my general relativity course,
unless you continue to derive general relativity by yourself!
\end{itemize}
\textsl{After exchanging pleasantries, Kim leaves for the long
cycle home.}

\textsl{Kim reflects that the thought experiment involving a
person in a windowless room who didn't know if the force they felt
was gravitational or centrifugal was very similar to the arguments
about absolute and relative motion that they learnt in their
special relativity course.}

\textsl{Sam contemplates this conversation with Kim and wonders if
it should be entered into next year's general relativity lecture
notes.}
\section*{Acknowledgements} The author would like to thank two
anonymous referees who provided insightful comments which improved
the clarity of arguments presented in this paper. The author would
also like to thank Vivienne Wheaton for proof reading the final
version.

\newpage
\appendix
\section{\label{app:clockGR}General Relativistic Corrections to the Effective Potential}
According to the general theory of relativity the proper time
interval ($d\tau$) for a clock in a weak gravitational field (such
as the earth's) is given by~\cite{schutz85}
\[-c^2d\tau^2 \approx
-\left(1+2\frac{\Phi_g}{c^2}\right)c^2dt^2 +
\left(1-2\frac{\Phi_g}{c^2}\right)dr^2 + +r^2d\theta^2 +
r^2\sin^2\theta d\phi^2 \quad , \] where $\Phi_g$ is given
by~Eq.\eqref{eq:spacePot} and $\frac{\Phi_g}{c^2} \ll 1$. For a
clock sitting on the surface of the earth
\[ dr = d\theta = 0 \text{ and } d\phi = \omega dt\]
so the proper time interval is
\begin{equation}\label{eq:dtau}
d\tau =  dt
\sqrt{1+\underbrace{2\frac{\Phi_g}{c^2}-\frac{r^2\omega^2}{c^2}\sin^2\theta}_{2\Phi_{ep}/c^2}}
\end{equation}
where $\Phi_{ep}$ is the newtonian gravitational potential
(see~Eq.\eqref{eq:effPot}). The time dilation effect is obtained by
rearranging~Eq.\eqref{eq:dtau}:
\begin{equation}\label{eq:a2_5}\frac{dt}{d\tau}  =
\frac{1}{\sqrt{1+2\frac{\Phi_{ep}}{c^2}}} \quad.
\end{equation}
We have shown in this paper that the weak equivalence principle
effectively states that time dilation can be calculated in terms of
the effective potential only, i.e.,
\begin{equation}\label{eq:a2_1}
\frac{dt}{d\tau}  = 1-\frac{\Phi_{ep}^{GR}}{c^2} \quad ,
\end{equation}
where $\Phi_{ep}^{GR}$ is the relativistic effective potential. The
relativistic effective potential can be determined in terms of the
newtonian potential by expanding~Eq.\eqref{eq:a2_5} and equating it
with~Eq.\eqref{eq:a2_1};
\begin{eqnarray}\label{eq:effPotG} \frac{\Phi_{ep}^{GR}}{c^2} & = & \frac{\Phi_{ep}}{c^2} -
\frac32\frac{\Phi^2_{ep}}{c^4} + {\cal
O}\left(\frac{\Phi_{ep}^3}{c^6}\right)  \nonumber\\
& \approx & \frac{\Phi_{ep}}{c^2} - \frac32\frac{\Phi^2_{ep}}{c^4}
\quad \text{; if } \frac{\Phi_{ep}}{c^2} \ll 1 \quad .
\end{eqnarray}
 Comparing~Eq.\eqref{eq:effPotG}
with~Eq.\eqref{eq:effPot}, using the values for $\Phi_{ep}$ as
calculated in~Eqs.\eqref{eq:effPotPole} and~\eqref{eq:effPotEq}  we
see that the relativistic effective potential differs from the
newtionian effective potential to a fractional accuracy of
$10^{-11}$.


\begin{thebibliography}{99}
\bibitem{colat} Colatitude is $pi/2$ minus latitude. It is the coordinate of choice for most references on planetary gravitational
potentials.
\bibitem{mass} This is the total mass of the Earth including the
    atmosphere. As we are calculating the gravitational potential at
    sea level we should subtract the mass of the atmosphere but we do
    not need to do this as the mass of the atmosphere is roughly one millionth of the total mass of the Earth.
\end{thebibliography}
\end{document}